\documentclass[10pt,
a4paper,
twocolumn
]{article}
\usepackage[eulergreek]{sansmath}
\usepackage[caption=false,position=top]{subfig}
\usepackage{array}
\usepackage{bm}
\usepackage[colorlinks=true,citecolor=blue,linkcolor=blue,
urlcolor=blue]{hyperref}
\usepackage{dcolumn}
\usepackage{mathtools}
\usepackage{amsmath}

\DeclarePairedDelimiterX\braket[2]{\langle}{\rangle}{#1 \delimsize\vert #2}

\DeclareMathAlphabet{\mathbfsf}{\encodingdefault}{\sfdefault}{bx}{n}

\newcolumntype{.}{D{.}{.}{-1}}

\newcommand{\ve}[1]{\mathbf{#1}}

\newcommand{\kf}[0]{k_\mathrm{F}}

\newcommand{\expe}[0]{\mathrm{e}}

\newcommand{\figref}[1]{Fig.~\ref{#1}}

\newcommand{\neweqnline}{\nonumber\\}

\newcommand{\eqnref}[1]{Eq.~(\ref{#1})}

\newcommand{\vecgrk}[1]{\boldsymbol{#1}}
\newcommand{\kfup}[0]{k_{\uparrow \mathrm{F}}}
\newcommand{\kfdn}[0]{k_{\downarrow \mathrm{F}}}
\newcommand{\kfsigma}[0]{k_{\sigma \mathrm{F}}}

\newcommand{\Nup}[0]{N_\uparrow}
\newcommand{\Ndn}[0]{N_\downarrow}

\newcommand{\Nsigma}[0]{N_\sigma}

\newcommand{\nuup}[0]{\nu_{\uparrow}}
\newcommand{\nudn}[0]{\nu_{\downarrow}}

\newcommand{\nusigma}[0]{\nu_{\sigma}}

\newcommand{\kD}[0]{k_\mathrm{D}}

\newcommand{\Ginv}[0]{\mathcal{G}^{-1}}
\newcommand{\Gg}[0]{\mathcal{G}}
\newcommand{\Jup}[0]{J_{\uparrow}}
\newcommand{\Jdn}[0]{J_{\downarrow}}
\newcommand{\Jupdn}[0]{J_{\uparrow \downarrow}}

\newcommand{\bfk}{\mathbf{k}}
\newcommand{\bfq}{\mathbf{q}}
\newcommand{\create}[1]{c^{\dagger}_{#1, \sigma}}
\newcommand{\crup}[1]{c^{\dagger}_{#1, \uparrow}}
\newcommand{\crdo}[1]{c^{\dagger}_{#1, \downarrow}}
\newcommand{\annihi}[1]{c^{}_{#1, \sigma}}
\newcommand{\anup}[1]{c^{}_{#1, \uparrow}}
\newcommand{\ando}[1]{c^{}_{#1, \downarrow}}
\begin{document}
\title{Communal pairing in spin-imbalanced Fermi gases}
\author{D.C.W.~Foo$^1$ \and T.M.~Whitehead$^1$ \and G.J.~Conduit$^1$}
\date{$^1$\emph{Cavendish Laboratory, J.J. Thomson Avenue, Cambridge, CB3 0HE, 
		United Kingdom}\\[2ex]}



\twocolumn[
\begin{@twocolumnfalse}
	\maketitle
	\begin{abstract}
A spin-imbalanced Fermi gas with an attractive contact interaction forms a superconducting state whose underlying components are superpositions of Cooper pairs that share minority-spin fermions.  This superconducting state includes correlations between all available fermions, making it energetically favorable to the Fulde--Ferrell--Larkin--Ovchinnikov superconducting state.  The ratio of the number of up- and down-spin fermions in the instability is set by the ratio of the up- and down-spin density of states in momentum at the Fermi surfaces, to fully utilize the accessible fermions.  We present analytical and complementary Diffusion Monte Carlo results for the state.
    \end{abstract}
\end{@twocolumnfalse}
]
\setlength\arraycolsep{1pt}

For over a century, the phenomenon of superconductivity has captured the attention of theorists, who have provided fundamental revelations about its underlying principles.  Bardeen, Cooper, and Schrieffer (BCS)~\cite{Bardeen57,Bardeen57a} gave seminal insights into the mechanism of superconductivity in systems with spin-balanced Fermi surfaces, showing that superconductivity may be understood as the collective behavior of coherent Cooper pairs of fermions.  Fulde, Ferrell, Larkin, and Ovchinnikov (FFLO)~\cite{Fulde64,Larkin65} extended this result, demonstrating that even in systems with spin-imbalanced Fermi surfaces Cooper pairs may still form the basis of a superconducting state.  However, in spin-imbalanced systems, the density of states in momentum at the Fermi surface of the majority-spin fermions is greater than that of the minority-spin fermions, so the number of Cooper pairs that can exist is limited by the number of minority-spin fermions at their Fermi surface.  This leaves many of the majority-spin fermions unpaired and so uncorrelated, wasting their potential for contributing correlation energy to the system.

\begin{figure}[t]
\centering
\includegraphics[width=0.45\linewidth]{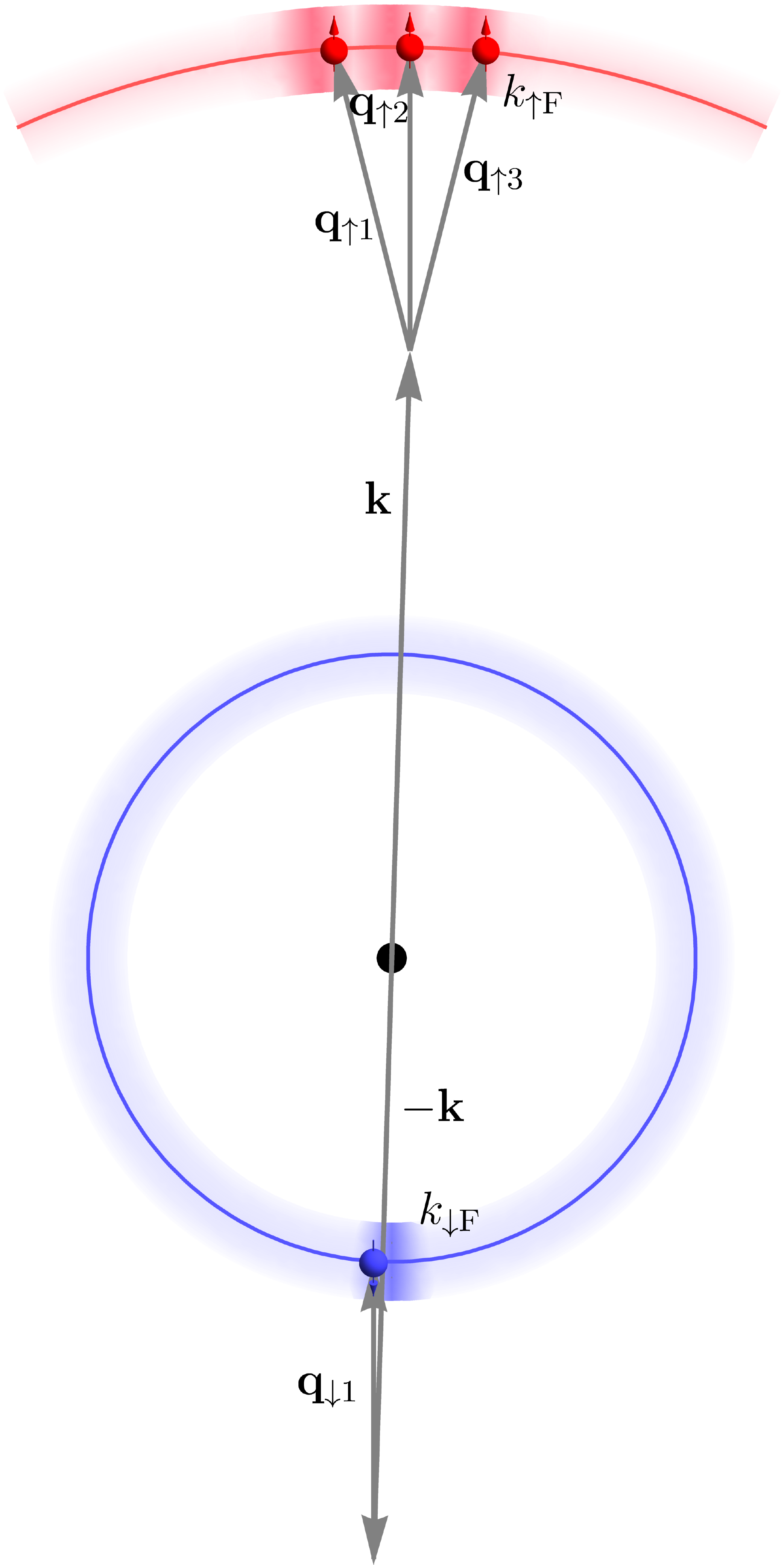}
\caption{(Color online) Idealized representation of the spin-imbalanced system showing Fermi surfaces for the down- (light blue circle) and up-spin (light red arc) species, with occupiable momentum states extending over a momentum scale set by the Debye frequency, forming annuli.  The intensity of color in the annuli indicate the approximate extent of the superconducting correlations.  Also shown are the momenta of $(\Nup,\Ndn)=(3,1)$ up- and down-spin fermions with corresponding $q$-vectors $\ve{q}_{\sigma i}$. The angular spread of the up-spin fermion momenta is exaggerated for clarity.}
\label{fig:ThreeToOne}
\end{figure}

In a few-fermion context, an instability~\cite{Whitehead17} containing more majority than minority-spin fermions maximizes the binding energy captured in spin-imbalanced systems by taking advantage of the correlations between all available momentum states.  Such an instability involves non-exclusive pairing between several majority-spin fermions and one minority-spin fermion in an ensemble that we call a communal state.  An example state is shown in \figref{fig:ThreeToOne}, with three majority- (up-)spin fermions each paired with the same minority- (down-)spin fermion.  This inspires us to merge Cooper pairs that share a minority-spin fermion to construct a communal superconducting state that correlates all available momentum states on the Fermi surfaces.  We show that this superconducting state with fermions shared between pairs is energetically favorable over the exclusive Cooper pair-based FFLO superconductivity in spin-imbalanced systems.

Current experimental developments enable the study of exotic superconducting phases in solid-state  spin-imbalanced Fermi gases~\cite{Bianchi03,Mayaffre14,Prozorov08,Canfield96,Hamidian16}, and spin-orbit coupling may give rise to inhomogeneous superconductivity~\cite{Zhang13,Lo14}; but no single experiment has provided unambiguous evidence for the existence of FFLO superconductivity, leaving the true nature of the ground state an open question.  However, the recent development of uniform trapping potentials for ultracold atomic gases~\cite{Mukherjee17} promises unparalleled experimental accuracy, presenting an ideal opportunity to revisit the structure of the superconducting ground state of spin-imbalanced Fermi gases.

In this Letter we examine the ratio of number of majority- to minority-spin fermions in communal states underlying the superconducting state of a spin-imbalanced Fermi gas, compare the energetics of communal superconductivity favorably to that of FFLO superconductivity, and present analytical and complementary Diffusion Monte Carlo results for the state.  We also discuss unique experimental consequences of the proposed communal superconductor.

To explore communal superconductivity we examine a two-spin fermionic system with an attractive contact interaction.  The quantum partition function, $\mathcal{Z}=\int\mathcal{D}(\psi,\bar{\psi})\expe^{-S[\psi,\bar{\psi}]}$, depends on the BCS action
\begin{align*}
S[\psi,\bar{\psi}]=\! &\sum_{\omega,\ve{k},\sigma}\!\bar{\psi}_{\ve{k},\sigma}\! (-i\omega\!+\!\xi_{\ve{k},\sigma})\psi_{\ve{k},\sigma} \neweqnline
&- g \!\!\!\sum_{\omega,\ve{k},\ve{k}',\ve{q}}\!\!\! \bar{\psi}_{\ve{k},\uparrow}\bar{\psi}_{\ve{q}-\ve{k},\downarrow}\psi_{\ve{q}-\ve{k}',\downarrow}\psi_{\ve{k}',\uparrow} ,
\end{align*}
where $\psi_{\ve{k},\sigma}$ and $\bar{\psi}_{\ve{k},\sigma}$ are a fermion field and its Grassmann conjugate, for momentum $\ve{k}$ and spin species \mbox{$\sigma\in\{\uparrow,\downarrow\}$}, $\xi_{\ve{k},\sigma} \equiv \epsilon_{\ve{k},\sigma}-\mu_\sigma$ where $\epsilon_{\ve{k},\sigma}$ and $\mu_\sigma$ are the species-dependent dispersion and chemical potential respectively, $g>0$ is the strength of the attractive contact interaction, and $\omega$ is a fermionic Matsubara frequency.  In this expression the momenta $\ve{q}$, referred to henceforth as $q$-vectors, give the net momenta of coupled fermions.  Our strategy is to build on the original BCS and FFLO theories that are directly applicable to the solid state, and so here adopt a Debye frequency cutoff on the sums over $\ve{k}$, however similar results are obtained in cold atom gases provided proper regularization is carried out.

We perform a Hubbard-Stratonovich decoupling in the Cooper channel, using a concise matrix formalism to express the action as
\begin{align}
S[\psi,\!\Delta]=&\sum_{\omega,\ve{k}}\begin{pmatrix}\bar{\vecgrk{\psi}}_\uparrow\\\vecgrk{\psi}_\downarrow\end{pmatrix}^\mathrm{T}\!\begin{pmatrix}\vecgrk{\Gg}^{-1}_\uparrow&&-\vecgrk{\Delta}\\-\vecgrk{\Delta}^\dagger&&\vecgrk{\Gg}^{-1}_\downarrow\end{pmatrix}\!\begin{pmatrix}\vecgrk{\psi}_\uparrow\\\bar{\vecgrk{\psi}}_\downarrow\end{pmatrix}\!\neweqnline&+\!\sum_\omega\!\frac{\mathrm{Tr}(\vecgrk{\Delta}^\dagger\!\vecgrk{\Delta})}{g},
\label{eq:action}
\end{align}
where the vectors $\vecgrk{\psi}_\sigma=(\psi_{(\ve{q}_{\sigma 1}+\varsigma_\sigma \ve{k}),\sigma},\psi_{(\ve{q}_{\sigma 2}+\varsigma_\sigma \ve{k}),\sigma},\ldots)^\mathrm{T}$, with \mbox{$\varsigma_\uparrow=+1$} and \mbox{$\varsigma_\downarrow=-1$}, the Grassman conjugates $\bar{\vecgrk{\psi}}_\sigma$ are similar, the matrices $\vecgrk{\Gg}^{-1}_\sigma=\mathrm{diag}(\Ginv_{(\ve{q}_{\sigma 1}+\varsigma_\sigma\ve{k}),\sigma},\Ginv_{(\ve{q}_{\sigma 2}+\varsigma_\sigma\ve{k}),\sigma},\ldots)$, for $\Ginv_{\ve{p},\sigma}=-i \omega+\varsigma_\sigma\xi_{\ve{p},\sigma}$, and
\begin{align*}
\vecgrk{\Delta}=\left(\begin{matrix}\Delta_{\ve{q}_{\uparrow 1}+\ve{q}_{\downarrow 1}} && \Delta_{\ve{q}_{\uparrow 1}+\ve{q}_{\downarrow 2}} && \cdots \\
\Delta_{\ve{q}_{\uparrow 2}+\ve{q}_{\downarrow 1}} && \Delta_{\ve{q}_{\uparrow 2}+\ve{q}_{\downarrow 2}} && \cdots \\
\vdots && \vdots && \ddots
\end{matrix}\right),
\end{align*}
where the $\ve{q}_{\sigma i}$ run over all the $q$-vectors of species $\sigma$.  We label the number of fermions in the underlying instability, and hence the number of $q$-vectors, per species by $\Nsigma$: therefore the $\vecgrk{\Gg}^{-1}_\sigma$ are $\Nsigma\times\Nsigma$ matrices and $\vecgrk{\Delta}$ is an $\Ndn\times\Nup$ matrix.  We shall find that in spin-imbalanced systems $\Nup\neq\Ndn$, so that $\vecgrk{\Delta}$ is rectangular rather than square, and different numbers of fermions from each species are involved in the underlying instability.  In the system represented by \figref{fig:ThreeToOne}, where there are three up-spin and one down-spin fermions involved in the underlying instability, $\vecgrk{\Delta}$ would be a $1\times 3$ matrix.  We focus our analysis on the Cooper channel as recent work~\cite{GMB2dpol} shows that in the BCS limit, screening~\cite{GMB64} and pairing mechanisms decouple so the reduction in critical temperature due to particle-hole interactions for both our communal state and for FFLO will be the same. We note for completeness that decoupling through the magnetic channel was also considered but had no consequence.

The elements of the $\vecgrk{\Delta}$ matrix gap the dispersion.  For the Fulde--Ferrell (FF) state~\cite{Fulde64} (also referred to as single-plane-wave superconductivity) $\vecgrk{\Delta}$ has only a single entry, and for crystalline FFLO superconductivity it is diagonal, see~\cite{Bowers02,Casalbuoni04,fflorevision2} and references therein.  The non-diagonal form here allows communal superconductivity, as in common with the few-fermion analysis~\cite{Whitehead17} multiple majority-spin fermions share a minority-spin fermion.  We focus on superconductivity where the sharing majority-spin fermions have nearly aligned $q$-vectors, comparable to the FF state.  For simplicity of analysis we assume that none of the $\ve{q}_{\uparrow i}+\ve{q}_{\downarrow j}$ pairs of $q$-vectors in $\vecgrk{\Delta}$ are degenerate.  Following Ref.~\cite{Whitehead17} the $\ve{q}_{\sigma i}$ vectors are taken to be not equal to each other so that each $\psi_{\ve{p},\sigma}$ appears only once in \eqnref{eq:action}.  This corresponds to assigning states on the Fermi surfaces into non-overlapping communal states of equal angular width.

With this expression for the action, working in the mean-field approximation, we can carry out a Ginzburg-Landau expansion of the regularized thermodynamic potential to obtain
\begin{align*}
\Omega=T\sum_{\omega,\ve{k}} \sum_{n=1}^\infty \frac{1}{n}\mathrm{Tr}(\vecgrk{\Gg}_\uparrow \vecgrk{\Delta} \vecgrk{\Gg}_\downarrow \vecgrk{\Delta}^\dagger)^n+\frac{\mathrm{Tr}(\vecgrk{\Delta}^\dagger\vecgrk{\Delta})}{g},
\end{align*}
where $T$ is the temperature.  To make progress with this expression, we symmetrize the coupling amplitudes, $\Delta_\ve{q}=\Delta$.  Near the second-order transition to the normal state we may neglect high-order terms in $\Delta$ and truncate the expression for the thermodynamic potential to
\begin{align*}
\Omega=\alpha \Delta^2 + \frac{1}{2}\beta \Delta^4 + \ldots,
\label{eq:Omega}
\end{align*}
where 
\begin{align}
\alpha&=\sum_{\ve{q}_\uparrow,\ve{q}_\downarrow}\left(\frac{1}{g} + T\sum_{\omega,\ve{k}}\Gg_{\ve{q}_{\uparrow}+\ve{k},\uparrow}\Gg_{\ve{q}_{\downarrow}-\ve{k},\downarrow}\right),\neweqnline
\beta&=\!\!\sum_{\substack{\ve{q}_{\uparrow 1},\ve{q}_{\downarrow 1},\\\ve{q}_{\uparrow 2}\,\ve{q}_{\downarrow 2}}}J(\ve{q}_{\uparrow 1},\ve{q}_{\downarrow 1},\ve{q}_{\uparrow 2},\ve{q}_{\downarrow 2}),
\end{align}
with
\begin{align}
J(\ve{q}_1,\ve{q}_2,&\ve{q}_3,\ve{q}_4)\!\neweqnline&=T\sum_{\omega,\ve{k}} \Gg_{\ve{q}_1+\ve{k},\uparrow}\Gg_{\ve{q}_2-\ve{k},\downarrow} \Gg_{\ve{q}_3+\ve{k},\uparrow}\Gg_{\ve{q}_4-\ve{k},\downarrow}.
\end{align}
To evaluate these expressions, we specialize to the case of small Debye frequency, found for many conventional superconductors~\cite{Kok37,Reynolds51,Horowitz52}.  In this limit, the vectors $\ve{q}_{\sigma i}$ are expected to be approximately parallel to maximise the number of contributing occupiable momentum states.  Approximately parallel but unequal $\ve{q}_{\sigma i}$ vectors provide a natural tiling of the Fermi surfaces into non-overlapping communal states.  This enables us to factorize out combinatorial factors, giving
\begin{align}
\alpha=\Nup\Ndn\Bigg(\frac{1}{g} &+ T\sum_{\omega,\ve{k}}\Gg_{\ve{q}+\ve{k},\uparrow}\Gg_{\ve{q}-\ve{k},\downarrow}\Bigg),\neweqnline
\beta=\Nup\Ndn \:\!\big[J_0 &+ (\Nup-1)\Jup+(\Ndn-1)\Jdn \neweqnline
&+(\Nup-1)(\Ndn-1)\Jupdn \big],
\label{eq:alphabeta}
\end{align}
where \mbox{$J_0=J(\ve{q},\ve{q},\ve{q},\ve{q})$}, \mbox{$\Jup=J(\ve{q}+\delta\ve{q}_\uparrow,\ve{q},\ve{q}-\delta\ve{q}_\uparrow,\ve{q})$}, \mbox{$\Jdn=J(\ve{q},\ve{q}+\delta\ve{q}_\downarrow,\ve{q},\ve{q}-\delta\ve{q}_\downarrow)$}, and $\Jupdn=J(\ve{q}+\delta\ve{q}_\uparrow,\ve{q}+\delta\ve{q}_\downarrow,\ve{q}-\delta\ve{q}_\uparrow,\ve{q}-\delta\ve{q}_\downarrow)$.  Here $\ve{q}$ is taken to represent the average $q$-vector for the fermions, symmetrized between species, and $\delta\ve{q}_\sigma$ is half the average separation between $q$-vectors for species $\sigma$, which in the small Debye frequency limit is orthogonal to the vector $\ve{q}$.  We follow the prescription of Ref.~\cite{Whitehead17} that the angular widths of the regions of Fermi surface involved in the communal superconducting state are the same between species, so the arc lengths $\delta\ve{q}_\sigma$ are proportional to the Fermi momenta and $|\delta\ve{q}_\uparrow|/|\delta\ve{q}_\downarrow|=\kfup/\kfdn$, where $\kfsigma$ is the Fermi momentum of species $\sigma$.  For a free dispersion $\Jupdn$ may be evaluated at zero temperature as

\begin{align}
\Jupdn = \frac{\mathcal{N}_d Q^2}{(Q^4\!-\!k_\perp^2(\delta\ve{q}_\uparrow\!+\!\delta\ve{q}_\downarrow)^2)(Q^4\!-\!k_\perp^2(\delta\ve{q}_\uparrow\!-\!\delta\ve{q}_\downarrow)^2)},
\label{eq:Jintegral}
\end{align}
%
%
where $\mathcal{N}_d$ is a dimension $d\in\{2,3\}$ dependent normalization factor, $Q^2\equiv2\kf\kD + \tfrac{1}{2}|\delta \ve{q}_\uparrow|^2 + \tfrac{1}{2}|\delta \ve{q}_\downarrow|^2 +\kD^2+k_\perp^2$, $\kf=(\kfup+\kfdn)/2$ is the average Fermi momentum, $\kD$ is the Debye frequency and $k_\perp$ is the average extent of $\ve{k}$ in the direction perpendicular to $\ve{q}$ such that the fermions are within the Debye frequency of the Fermi energy.  Similar expressions for $\Jup$, $\Jdn$ and $J_0$ can be found by taking $\delta\ve{q}_\downarrow=0$, $\delta\ve{q}_\uparrow=0$, or both.  For $\kD\ll\kf$, $k_\perp\sim\sqrt{\kf\kD}$ and \eqnref{eq:Jintegral} confirms that for a single instability $\beta\ge 0$ for realistic values of the $\ve{q}_i$~\cite{Casalbuoni04}, justifying the truncation in \eqnref{eq:Omega}.

To identify the optimal ratio of number of fermions involved in the communal superconductor, we express \eqnref{eq:Omega} as a function of $\Nup/\Ndn$ and $\Nup\Ndn$, and then perform a saddle point analysis to optimize $\Omega$ with respect to $\Nup/\Ndn$, $\Nup\Ndn$, and $\Delta$ simultaneously. We note from Ref.~\cite{Whitehead17} that singly-excited state fluctuations in $\Nup/\Ndn$ have a linear energy dependence, so we focus on this saddle point analysis.  This gives the expected ratio of number of fermions involved in the underlying instability as
\begin{align}
\frac{\Nup}{\Ndn}=\frac{\Jupdn-\Jdn}{\Jupdn-\Jup}=\left(\frac{|\delta\ve{q}_\uparrow|}{|\delta\ve{q}_\downarrow|}\right)^2=\left(\frac{\nuup}{\nudn}\right)^{2/(d-1)},
\label{eq:NupNdn1}
\end{align}
where the second equality was obtained from \eqnref{eq:Jintegral} and $\nusigma$ is the density of states in momentum at the Fermi surface of species $\sigma$.

In the case of cold atoms in the absence of a Debye frequency, regularization of the divergent momentum summation in $\alpha$ can be done by using scattering theory to replace the weak interaction coupling parameter $g$ with a formal expression involving the scattering length and another summation with the same UV divergence characteristics, resulting in a convergent expression~\cite{fermigasreview}. The momentum summation in $\beta$, and indeed in all higher terms of the expansion of $\Omega$, do not exhibit UV divergences, as can be seen in how the expression for $\Jupdn$ in~\eqnref{eq:Jintegral} is finite for any value of $\kD$, indicating that the high $\ve{k}$ contributions to the summation are not dominant. Indeed, provided the interaction is weak, the optimal placement of the $\ve{q}$ are unchanged and since the dominant contributions to the $J$ come from around the Fermi surface, the result of~\eqnref{eq:NupNdn1} remains unchanged, that is making the shift from solid-state to ultracold atoms does not change any of the significant physics.

This result confirms that the superconducting state is indeed communal, with pairs sharing minority-spin fermions to take advantage of all available correlations in spin-imbalanced systems. \eqnref{eq:NupNdn1} also aligns with our heuristic expectation that the instability involves more fermions of the species with the larger density of states in momentum at its Fermi surface, as was also found in the few-fermion case~\cite{Whitehead17}.  For spin-balanced systems, $\nuup=\nudn$ and so $\Nup/\Ndn=1$, recovering the BCS theory result, whilst in the polaron limit of a single minority-spin impurity in a full Fermi sea of majority-spin fermions, the single minority-spin fermion couples with all the majority-spin fermions at their Fermi surface, in agreement with results from the literature~\cite{Chevy06,Combescot08,Massignan14}.  

Our conclusions do not contradict the well-known result that FFLO superconductivity is the ground state in one dimension, obtained separately by density matrix renormalization group~\cite{fflo1ddmrg} and time-evolving block decimation methods~\cite{fflo1dtebd}. In a one-dimensional system $\nuup=\nudn$ regardless of the spin-imbalance, removing our initial physical motivation for communal pairing. Furthermore, in one dimension $\delta\ve{q}_\sigma$ cannot be orthogonal to $\ve{q}$, and so if one attempts to form a communal state, it necessarily involves fermions of a particular spin species with different energies, invalidating the symmetrization of the gap amplitudes $\Delta_\ve{q}$ so there is no evidence that communal pairing can be energetically favourable to the FFLO state found in previous studies of one-dimensional systems.

The same optimization procedure that gave \eqnref{eq:NupNdn1} for $\Nup/\Ndn$ also provides an expression for $\Nup\Ndn$, as
\begin{align}
\Nup\Ndn=\frac{\left(J_0-\Jup-\Jdn+\Jupdn\right)^2}{\left(\Jupdn-\Jup\right)\left(\Jupdn-\Jdn\right)}.
\label{eq:NupNdn3}
\end{align}
For reasonable values of the $|\delta\ve{q}_\sigma|$ and $|\ve{q}|$ in spin-imbalanced systems this expression gives values of $\Nup\Ndn>1$, confirming that the communal superconductor is indeed made up of multiple fermions of at least one spin species.  Excessively high $\Nup\Ndn$ is energetically penalized by the highest term in the expansion of the thermodynamic potential, which goes as $(\Nup\Ndn\Delta^2)^n$, and so we expect communal superconductivity to have both $\Nsigma$ being reasonably small integers.  In the spin-balanced limit \eqnref{eq:NupNdn3} collapses to the BCS result $\Nup\Ndn=1$.

The analysis may be adapted to the number conserving canonical ensemble by constructing the Helmholtz free energy $F = \Omega + \mu_\uparrow n_\uparrow + \mu_\downarrow n_\downarrow$, where $n_\sigma = - \partial\Omega/\partial\mu_\sigma$ is the total particle number of spin species $\sigma$, so $F = (1-\mu_\uparrow\partial/\partial\mu_\uparrow-\mu_\downarrow\partial/\partial\mu_\downarrow)\Omega$.  As the chemical potential only appears in the propagator, it suffices to note that $\partial \Gg_{\ve{k},\sigma}/\partial\mu_\nu = \varsigma_\sigma\Gg^2_{\ve{k},\sigma}\delta_{\sigma,\nu}$ and so the net result on $\Nup\Ndn$ and $\Nup/\Ndn$ of moving from the grand canonical ensemble to the canonical ensemble is to shift the functions $J$ as 

\begin{align*}
J\to J-&T\sum_{\omega,\ve{k}} \Gg_{\ve{q}_1+\ve{k},\uparrow}\Gg_{\ve{q}_2-\ve{k},\downarrow} \Gg_{\ve{q}_3+\ve{k},\uparrow}\Gg_{\ve{q}_4-\ve{k},\downarrow}
\neweqnline
&\times(\mu_\uparrow(\Gg_{\ve{q}_1+\ve{k},\uparrow}+\Gg_{\ve{q}_3+\ve{k},\uparrow})\neweqnline&\quad-\mu_\downarrow(\Gg_{\ve{q}_2-\ve{k},\downarrow}+\Gg_{\ve{q}_4-\ve{k},\downarrow})),
\label{Jshift}
\end{align*}
leading to the same conclusions as in the grand canonical ensemble, namely that in the canonical ensemble the superconducting state is indeed communal with $\Nup\Ndn > 1$ and $\Nup/\Ndn = (\nu_\uparrow/\nu_\downarrow)^{(2/(d-1))}$.

Now that we have shown that the communal superconductor is energetically favourable over single-plane-wave FFLO superconductivity in a spin-imbalanced Fermi gas, we need to confirm that we have not compromised the stability of the superconducting state.  We validate this by examining the phase boundaries between the communal superconductor and three competitor phases.

With increasing spin-imbalance, BCS superconductivity becomes unstable against FFLO superconductivity at the Chandrasekhar-Clogston limit~\cite{Chandrasekhar62,Clogston62}.  Although communal superconductivity is energetically favourable over FFLO superconductivity, BCS superconductivity still has a large density-of-states advantage over communal superconductivity, and to a first approximation the phase boundary between BCS superconductivity and communal superconductivity will remain at the same Chandrasekhar-Clogston value.

The phase boundary between communal superconductivity and the normal state will also remain the same as between single-plane-wave FFLO superconductivity and the normal state.  In both cases the second-order phase transition occurs when $\alpha=0$, and this condition is identical between FFLO and communal superconductivity, up to an irrelevant multiplicative factor of $\Nup\Ndn$ in \eqnref{eq:alphabeta}, and so the phase boundary is also identical.

Stability against phase separation can be expressed as the positive-definiteness of the total particle number susceptibility matrix~\cite{Chen06,Wang17}.  This condition includes the possibility of separation into two superconducting phases, with ratios of number of fermions differing from that predicted in \eqnref{eq:NupNdn1}, and may be expanded following \eqnref{eq:Omega}, to leading order in $\Delta$ giving

\begin{align*}
 \alpha\frac{\partial^2\alpha}{\partial q^2} > 2\left(\frac{\partial\alpha}{\partial q}\right)^2.
\end{align*}

This is the same as the equivalent expression for FFLO superconductivity, up to a factor of $(\Nup\Ndn)^2$ that cancels between the two sides of the inequality.  This indicates that the line of stability against phase separation is the same for communal superconductivity as for FFLO superconductivity to leading order.

Although the discussion above focuses on nearly aligned $q$-vectors, comparable to the FF state, it is known that the Larkin--Ovchinnikov (LO) state~\cite{Larkin65} built from two plane-waves can be energetically favorable to single-plane-wave superconductivity.  Therefore, we now follow the prescription of Larkin and Ovchinnikov and consider a communal superconducting state out of two instabilities on opposite sides of the Fermi surfaces.  The only differences in the theory of communal superconductivity for one and two instabilities are a multiplicative factor of 2 in \eqnref{eq:Omega} and additional terms in the expression for $\beta$ in \eqnref{eq:alphabeta}.  Similarly to the single instability case the optimal instability contains more up- than down-spin fermions, and so the communal superconducting state is also energetically favorable over the LO state.

\begin{figure}[t]
\centering
	\includegraphics[width=0.9\linewidth]{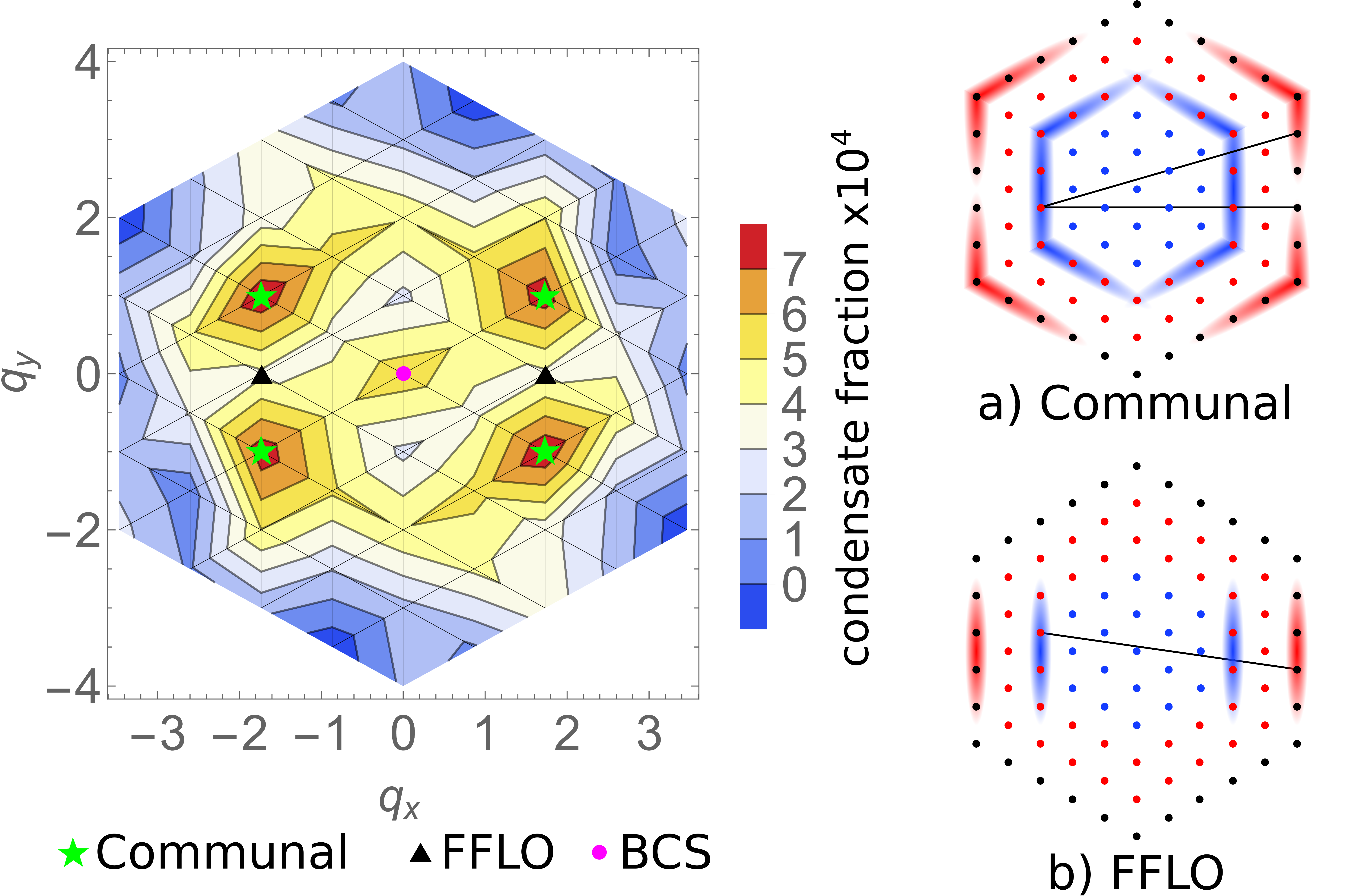}
	\caption{(Color online) Left: Plot of condensate fraction in pair momentum space in units of $1/2\sqrt{3}\pi r_s$.  The positions of expected peaks corresponding to a communal state ($r_s q=1/\sqrt{3}\pi$), FFLO pairing ($r_s q=1/2\pi$), and BCS pairing ($r_s q=0$) are shown with green stars, black triangles and a magenta circle respectively.  Communal type peaks are prominent with no obvious FFLO peaks and a mild BCS peak.  The k-space grid is drawn in black.  Right: Diagrams of pairing orbitals in the communal and FFLO phases.  The colored points indicate the filled Fermi areas of the down-spin (blue) and up-spin (red) species in the non-interacting limit.  The colored clouds indicate which states contribute to the instabilities with intensity indicating the strength of contribution.  Black lines indicate a possible pairing arrangement.}
	\label{fig:DMCdata}
\end{figure}

We supplement the preceding analysis with numerical evidence obtained from a quantum Monte Carlo study of a finite spin-imbalanced 2D homogeneous fermion gas with attractive interactions using the \textsc{casino} program~\cite{casino}.  To minimize finite size effects~\cite{finitesizeeffect,simdark}, we place the fermions in a rhomboidal box with vertex angle $60^\circ$ so that the discretized momentum points form a triangular lattice.  This allows for the densest tiling of discrete momentum points in 2D giving the closest to circular Fermi surfaces. We set up a system of 61 spin-up and 19 spin-down fermions and work in atomic units with the average inter-fermion separation $r_s=1$.  Such particle numbers are consistent with those used in other DMC studies~\cite{dmcN1,dmcN2}, giving us confidence that the results obtained should be at least qualitatively related to the analytics done in the thermodynamic limit.  We note for completeness that qualitatively similar results were obtained for systems with both a smaller and larger number of fermions, and in systems on a square simulation cell.  

As the Fermi surfaces form hexagons and have a 2:1 ratio of fermions, we expect a $(\Nup,\Ndn) = (2,1)$ instability.  An ultratransferable pseudopotential~\cite{pseudopot} with scattering length $a=6.5991$ and zero effective range was introduced so that the BCS coherence length of an equivalent spin-balanced system was approximately equal to the simulation cell size.  We note here for completeness that while such a pseudopotential is indeed meant to emulate the scattering properties of a contact interaction, it nevertheless has a finite extent in space that introduces a natural momentum cutoff.  The UTP therefore does not have infinite range in momentum space and so the assumption of small Debye frequency used in the analytical derivation is applicable to the simulation.

Following previous work~\cite{riospairingorb}, we employ a Slater-Jastrow trial wavefunction of the form $\Psi_\mathrm{T} = \mathrm{e}^{-J(\mathsf{r_\uparrow},\mathsf{r_\downarrow})}\det[\phi(\ve{s}_{i,j})].$ The pairing orbital is

\begin{align*}
\phi(\ve{s}_{i,j}) = &\sum_{l=1}^{4}a_l \cos (\ve{k}_l\cdot\ve{s}_{i,j}) \neweqnline &+ \Theta(\Ndn-i)(1-\tfrac{s}{r_s})^3\Theta(1 -\tfrac{s}{r_s})\sum_{m=0}^{2}b_m r^m,
\end{align*}
where $\ve{s}_{i,j} \equiv \ve{r}_{\uparrow,i} - \Theta(\Ndn-i)\ve{r}_{\downarrow,j}$, $s\equiv\lvert\ve{s}_{i,j}\rvert$, $\ve{r}_{\sigma,i}$ is the position vector of the $i-$th fermion of spin species $\sigma$, $\ve{k}_l$ is the l$th$ shortest reciprocal-space vector, $\Theta$ is the Heaviside step function, and the $\{a_l\},\{b_m\}$ are optimisable parameters. The Jastrow factor is given by

\begin{align*}
 J(\mathsf{r_\uparrow},\mathsf{r_\downarrow}) = \sum_{i,j}\Big[ &\sum_k u_k r_{i,j}^{k-1}(1-\frac{r_{i,j}}{r_s})^3\Theta(1-\frac{r_{i,j}}{r_s}) \\ + &\sum_m p_m \cos(\ve{G}_m\cdot\ve{r}_{i,j})+ \nu\ \mathrm{ terms} \Big]
\end{align*}
where $\mathsf{r_\sigma}$ denotes the set of position vectors for all fermions of spin species $\sigma$, $\ve{r}_{i,j}\equiv\ve{r}_{\uparrow,i}-\ve{r}_{\downarrow,j}$, $r\equiv\lvert\ve{r}_{i,j}\rvert$, the $\ve{G}_m$ are reciprocal-space vectors through which anisotropy may be introduced, and the $\{u_k\},\{p_m\}$ are optimisable parameters. $J(\mathsf{r_\uparrow},\mathsf{r_\downarrow})$ is thus a function of all opposite-spin fermion separations containing a short range isotropic $u$ term, anisotropic $p$ terms~\cite{drumjas} and a $\nu$ term~\cite{nujas} that reflects the simulation cell symmetry and whose form is omitted for brevity. While the trial wavefunction was originally used to capture pairing between electrons and holes in a bilayer, it nevertheless has three attractive properties that warrant usage in this context; namely that it is a pairing wavefunction and that it deforms continuously into the accepted form for a balanced superconductor and the non-interacting state. Furthermore, the nodal surface can vary and be optimized through the $b_m$ parameters.

The trial wavefunction was optimised using Variational Monte Carlo~\cite{vmcopt} then equilibrated using Diffusion Monte Carlo before the condensate fraction in momentum space~\cite{condfracq} was accumulated. In \textsc{casino}, the condensate fraction is defined as $f_{\bfq} \equiv \sum_\bfk (\langle\crup{\bfk}\crdo{\bfq-\bfk}\ando{\bfq-\bfk}\anup{\bfk}\rangle - n_{\bfk,\uparrow}n_{\bfq-\bfk,\downarrow})$, where $\create{\bfk} (\annihi{\bfk})$ is the creation (annihilation) operator for a fermion of momentum $\bfk$ and spin $\sigma$ and $n_{\bfk,\sigma} \equiv \langle\create{\bfk}\annihi{\bfk}\rangle$ is the momentum density. The results display strong anisotropy of the condensate fraction at $r_s q=1/\sqrt{3}\pi$, regardless of whether in the Jastrow factor the $p_m$ are selected to permit anisotropy, constrained to ensure isotropy, or set to zero. The state observed is therefore not an artefect of (an)isotropy of the trial wavefunction, consistent with the communal state being the robust ground state. Furthermore, we confirm the sharing of minority-spin fermions between pairs by noting that the anisotropy is consistent with the pairing scheme shown in the upper right of Fig.~\ref{fig:DMCdata}; the black lines indicate the up-down pairs present in a single $(\Nup,\Ndn) = (2,1)$ instability which when combined with its time-reversed partner gives 4 pairing momenta, corresponding to 4 peaks in the condensate fraction. We only indicate the peak positions for plane wave FFLO; crystalline FFLO would have peaks at all $q$-vectors of equal magnitude to plane-wave FFLO, none of which coincide with the 4 observed peaks of the communal pairing state.

The lack of a strong condensate fraction at $r_s q=1/2\pi$ indicates that the FFLO phase is not the major contributor to the ground state. Finally we note for completeness that while there appears to be a BCS contribution to the condensate fraction, the uncertainty in the condensate fraction at $r_s q=0$ was twice as large as at other points and so that peak is half as significant as the others. Similar results were seen in systems with a 3:1 and 4:1 ratio at the Fermi surface, affirming the pairing result for $\Nup/\Ndn$, and for systems with the same 2:1 ratio at the Fermi surfaces but different total numbers of fermions, namely 19 spin-up and 7 spin-down fermions, and 127 spin-up and 37 spin-down fermions. The results obtained for systems with different numbers of fermions give us confidence that the observed state is not an artefact of finite size effects.

We note for completeness that the condensate fraction was also gathered for pairs of the same spin species in an attempt to simultaneously search for induced p-wave superfluidity~\cite{fflorevision3}, however the values of intra-spin condensate fraction were more than 10 orders of magnitude smaller than those for the inter-spin condensate fraction and were indistinguishable from zero at all pair momenta.

Having seen that communal superconductivity is energetically favorable over FFLO superconductivity, we now consider its possible experimental consequences.  We focus on two effects where the communal nature of the underlying instability should be directly observable.

\begin{figure}[t]
\centering
	\includegraphics[width=1\linewidth]{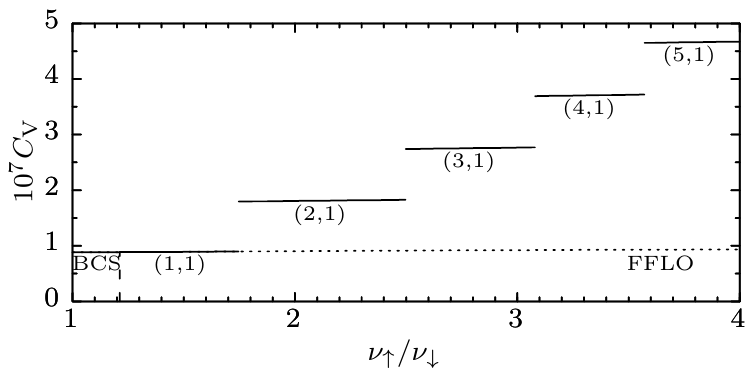}
	\caption{Plot of specific heat $C_{\mathrm{V}}$ against $\nuup/\nudn$ for the energetically favourable communal (solid) and not energetically favourable FFLO (dotted) phases. Communal phases are labelled by the pair $(\Nup,\Ndn)$. The BCS phase denotes where the Chandrasekhar-Clogston limit is breached, indicated by the vertical dashed line. The two curves coincide in the BCS and (1,1) phases.}
	\label{fig:CVdata}
\end{figure}

\textbf{Multiple phase transitions:}  With increasing spin-imbalance, that is increasing $\nuup/\nudn$, \eqnref{eq:NupNdn1} predicts that the ratio $\Nup/\Ndn$ should increase.  Starting from the BCS state with $\nuup/\nudn=1$ the system should progress through several communal superconducting phases with increasing values of $\Nup/\Ndn$, where both $\Nsigma$ are integers, giving a series of different superconducting states.  The product $\Nup\Ndn$ governed by \eqnref{eq:NupNdn3} restricts $\Nup$ and $\Ndn$ to small integers and so restricts the number of transitions observed.  Each transition is expected to be second order, and so the communal superconducting phase would be characterized by a series of discontinuities in the heat capacity and the compressibility, which is directly observable in ultracold atomic gases~\cite{Ku12}, as the spin-imbalance is changed. No such phase transitions are expected within the FFLO phase. Plots of the specific heat $C_{\mathrm{V}}$ against $\nuup/\nudn$ at low temperature are shown in~\figref{fig:CVdata} for the communal and FFLO phases. The plot used the same parameters as our DMC study. A finite discontinuity in the specific heat is present in both curves at the Chandrasekhar-Clogston limit at $\nuup/\nudn\approx1.2$, with further discontinuities manifesting in the communal phase as $\Nup$ and $\Ndn$ change. With increasing imbalance the heat capacity jumps upward as an additional up-spin fermion enters the strongly correlated state, increasing its binding energy. The behavior of the communal state is in marked contrast to the FFLO phase that has near constant specific heat, giving a significant observable difference
between the two phases.

\textbf{Superconducting order parameter:} In real space the order parameter will exhibit a beat pattern due to the interference between similar $q$-vectors, which could allow the identification of the particular $q$-vectors in the superconductor.  The order parameter and its spread in momentum could be determined in an ultracold atomic gas experiment through density-density correlations measured from time-of-flight experiments~\cite{Altman04}. While the width of the narrow peaks of either the BCS or FFLO phase can be resolved in absorption imaging experiments to better than $0.1\kf$~\cite{cfracspreadexpt}, the spread of $q$-vectors from communal pairing seen in~\figref{fig:DMCdata} and indicated by exact diagonalisation studies~\cite{Whitehead17} is on the order of $\kf$ and should be readily distinguished. This can be seen in Fig.~\ref{fig:DMCdata} where alongside the clear peaks, the condensate fraction is generally nonzero for momenta $r_s q \leq 1/\sqrt{3\pi}$ and only falls to zero on the border of the region shown. In contrast, FFLO and crystalline FFLO theories predict sharp peaks in the condensate fraction, as in spin-balanced BCS theory, at fixed magnitude of the pairing momenta.

\textbf{Andreev reflection:} The elementary excitations above the proposed ground state are predicted to be well-described by the few fermion analysis~\cite{Whitehead17}. This should have novel consequences especially concerning Andreev reflection experiments as the strong correlations between a group of fermions held in a communal state should result in multiple retroreflected holes for a single incident fermion, in sharp contrast to the single hole per fermion expected in normal FFLO theory.

Beyond these experimental signatures, communal superconductivity also introduces the notion that the building block of a superconductor may involve the sharing of fermions between Cooper pairs.  In particular, there can be fluctuations in the number of fermions in the underlying instability, which could lead to the renormalization of the properties of a spin-balanced superconductor.   The analysis is also generalizable to systems with multiple underlying instabilities, more akin to crystalline FFLO superconductivity~\cite{Bowers02}, and we use that system as a guide for the likely modifications when communal superconductivity is built from several instabilities.

We have introduced the idea of a communal superconductor, whose underlying instability is composed of multiple pairs with shared fermions, to enable the use of all available inter-fermion correlations.  We have shown that communal superconductivity is energetically favorable over FFLO superconductivity in spin-imbalanced Fermi gases, both analytically and with complementary DMC results, and that a communal superconductor has clear experimental signatures.

Data used for this Letter are available online~\cite{FigData}.
  The authors thank Adam Nahum, Johannes Hofmann, Johannes Knolle, Jens Paaske, 
Robin Reuvers, and Victor Jouffrey for useful
  discussions, and acknowledge the financial support of the NUS, the EPSRC, and the Royal 
Society.

\end{document}